# Possible excited spin density wave states in zigzag graphene nanoribbons

Hang Xie[1,*] Jinhua Gao[2] and Dezhuan Han[1]

[1] School of Physics, Chongqing University
[2] School of Physics, Huazhong University of Science and Technology

We theoretically examine the possible spin ordered states in zigzag graphene nanoribbon in a large supercell by the self-consistent mean field method as well as the first principle calculation. In addition to the well-known anti-ferromagnetic (AF) and ferromagnetic (FM) edge states, we find that there are also some excited spin density wave (ESDW) states, the energy of which are close to the AF edge state (ground state). We thus argue that these ESDW states may coexist in experiment when the temperature is not too low. Our numerical results indicate that the allowed ESDW are commensurate; and their dispersion curve is plotted. This interesting feature results from the fact the there is a flat edge band near the Fermi surface in the graphene zigzag ribbon. Finally we use the non-equilibrium Green's function theory combined with the Hubbard model to calculate some possible spin polarization states localized in the middle and terminal parts of graphene nanoribbons.

## I. INTRODUCTION

Graphene, an important two-dimensional material, has attracted a lot of research interests in the past decade [1-3]. It has many novel properties such as Dirac-like electron behavior with the Klein tunneling effect [2, 4]; the high opacity in the optical wavelength range [5]; the edge states on the zigzag graphene nanoribbons (GNR) [6]. For the GNRs, they have different energy gaps due to the widths and axis orientations and they may be metallic, semiconducting or insulate [7].

Among them, the magnetic properties of graphene are very interesting. The small spin-orbital interaction and the long spin diffusion length lead graphene as an ideal spintronics material. It is well known that there exist spontaneous spin polarizations on the edges of zigzag GNR [8]. That means the two ribbon-edge regions have either anti-ferromagnetic (AF) or ferromagnetic (FM) states: the later are often induced by the external magnetic field. The FM state often has a little higher energy than the AF state. The energy difference decays as the increase of the ribbon width [8, 9]. Based on the first-principle calculations or the tight-binding method with Hubbard model, many people proposed some spintronic structures, such as the anti-dot lattices [10], the thaw-tooth GNRs [11]; the porous graphene sheets doped with nitrogen and boron atoms [12] or the dumb-bell nanoislands [13]. The energy bands of these materials often exhibit half-metallic property with potential applications in the spintronic devices. Besides these periodic graphene systems, some people also found that there exist magnetic orders in finite graphene nanoislands [14]. These ground states are in agreement with the Lieb's theorem [15].

However, most of these magnetic orders, existing on the zigzag edges, are either in the same direction (FM order) or the opposite direction (AF order) [16]. A very few works show more complicated magnetic states on one zigzag edge. K.W. Kim and A.K. Kim's calculations show that on the graphene-based spin valve devices, the zigzag GNR's magnetism may have opposite FM states with a single domain wall (DM) in the middle part due to the different magnetic field in the leads [17]. With the first principle calculations and NEGF theory, they also obtained the transmission currents with some bias voltages. Furthermore, the spin waves in zigzag GNR have been studied with the spin transverse dynamic susceptibility theory [18-21]. The dispersion relation and the lifetime of these spin excitations are investigated. However, these studies are mainly focused in the susceptibility spectra. The investigation is limited to the model system and simple excitation modes; the spin polarization distribution has seldom been seen. We also notice that someone proposed that there exist spin density waves in the periodically strained GNR by the first principle calculations [22]. On the experiment aspect, G.Z. Magda et al observed a semiconductor-to-metal transition in the GNR measurements at the room temperature, which is due to the magnetic order change from the AF state to the FM state [23].

In this paper we report that there exist some intrinsic complicated spin excitations in a uniform zigzag GNR. These modes, corresponding to some excited spin-density-wave (ESDW), have very similar band gaps as insulators. Their energies are close to the AF or FM state. So they can be measured in the experiments as well. These states consist of various FM or AF orders or their hybrid combinations. The calculation is based on the tight-binding and Hubbard models with the self-consistent calculations, as well as the first principle calculations. Combined with the non-equilibrium Green's function (NEGF) theory, we find that in the local region of zigzag GNR, there also exist similar SDW excitations. All these novel states can be viewed as some spin excitations in the configuration space of magnetic orders. Some energy analysis and band structures have been employed for these states.

This paper is organized as follows: in Sec. II we give an introduction to the relative theories, including the Hubbard model, the band structure calculation and the NEGF theory; in Sec. III we show the results and discussions for the novel ESDW states in the periodic and non-periodic GNR systems. In Sec. IV we draw our main conclusions. The contour integral for the density matrix calculation is given in the appendix.

## II. THEORY
### A. Hubbard model and band structure calculation

In the unrestricted Hartree-Fock (UHF) model with the mean-field approximation [24-26], the Hamiltonian is written as

$$H = \sum_{<i,j>} t c_{i\uparrow}^\dagger c_{j\uparrow} + \sum_{<i,j>} t c_{i\downarrow}^\dagger c_{j\downarrow} + U \cdot \sum_{i}^{N_0} [n_{i\uparrow} <n_{i\downarrow}> + n_{i\downarrow} <n_{i\uparrow}>] \tag{1}$$

The first two terms are the hopping energy of the tight-binding (TB) model with two spin components. $<i,j>$ means the hopping terms exist only between the nearest neighboring sites and $t=-2.7$ eV. The last term is also called the Hubbard term, resulting from the Coulomb repulsion of two on-site electrons with different spins. Here we choose $U=2.0$ eV for this value can best reproduce the first-principle calculation results for zigzag GNR [25-26]. $<n_{i\downarrow}>$ and $<n_{i\uparrow}>$ are the average electron densities at site $i$ with different spins.

In a periodic system the band structure is obtained from the eigenvalue calculation [27],

$$\sum_{j}^{N_0} [\sum_{n} H_{i_0, j+n \cdot N_0} \cdot e^{i\mathbf{k} \cdot \mathbf{R}_n}] a_j = E \sum_{j}^{N_0} \delta_{i_0, j} a_j , \tag{2}$$

where $N_0$ is the number of atoms in a unit cell, $k$ is the Bloch wavevector, $n=-1,0,1$, corresponds to three neighboring unit cells (in 1D case). $\delta_{i,j} = 1$ $(i = j)$, or $0$ $(i \neq j)$, comes from the orthogonal relation of two basis functions in the tight-binding model. Eq. (2) may be rewritten in a matrix form

$$\mathbf{H}(k)\mathbf{a}_p = E_p \mathbf{I} \mathbf{a}_p. \qquad (3)$$

where $\mathbf{H}(k)$ is from the combination of the Hamiltonians (in Eq. (1)) in the neighboring unit cells with the weight factors of $e^{i\mathbf{k}\cdot\mathbf{R}_n}$.

When considering the Hubbard term, we may double the dimension of the Hamiltonian and the eigen-basis with two spin components. Instead of doubling the dimension, here we use two sets of eigenvalue equations corresponding to two spins. From Eq. (1) the diagonal elements of the each Hamiltonian consists of the electron densities with opposite spin, which makes the two equations coupled to each other. This is a self-consistent problem and has to be solved with the numerical iterations. In Eq. (1), $<n_i^{\uparrow(\downarrow)}>$ is the mean electron density on each site, which comes from all eigenstates in the occupied orbitals and averaged in the first Brillouin zone, as shown below [25]

$$<n_i^{\uparrow(\downarrow)}> = \frac{1}{(\pi/R)}\int_0^{\pi/R} dK \cdot n_i^{\uparrow(\downarrow)}(K). \qquad (4)$$

where

$$n_i^{\uparrow(\downarrow)}(K) = \sum_p^{occ} |A_{i,p}^{\uparrow(\downarrow)}(K)|^2, \qquad (5)$$

'*occ*' means the summation is utilized only for the occupied orbitals; $\pi/R$ is the volume of the first Brillouin zone (1D case). $A_{i,p}^{\uparrow(\downarrow)}(K)$ is the eigenvector of $i^{th}$ site for the $p^{th}$ eigenvalue with the Bloch wavevector $K$.

### B. NEGF theory

Besides the periodic GNR, we also investigate the non-periodic GNR with the none-equilibrium Green's function (NEGF) theory [28]. The Hubbard term exist both in the device and the leads.

$$[E\mathbf{I} - \mathbf{H}(<n^{\downarrow(\uparrow)}>) - \mathbf{\Sigma}^r(E, <n^{\downarrow(\uparrow)}>)] \cdot \mathbf{G}^{r,\uparrow(\downarrow)}(E) = \mathbf{I}, \qquad (6)$$

$$\mathbf{\Sigma}_\alpha^r(E, <n^{\downarrow(\uparrow)}>) = \mathbf{h}_{D\alpha} \cdot \mathbf{g}_\alpha^r(E, <n^{\downarrow(\uparrow)}>) \cdot \mathbf{h}_{\alpha D}, \qquad (7)$$

where $\mathbf{I}$ is the unitary matrix with the dimension of the device; $\mathbf{\Sigma}^r = \mathbf{\Sigma}_L^r + \mathbf{\Sigma}_R^r$ is the retarded self-energy, which is given by the coupling Hamiltonians and the surface Green's function $\mathbf{g}_\alpha^r$

(see Eq. (7), $\alpha = L, R$, which means the left or right lead.) $\mathbf{g}_\alpha^r$ is obtained from the recursive Green's function method [29]. The imaginary part of $G_{i,i}^{r,\uparrow(\downarrow)}$ gives the local density of states (LDOS): $\rho_i^{\uparrow(\downarrow)}$, which is then integrated to obtain the electron density $<n_i^{\downarrow(\uparrow)}>$

$$\rho_i^{\uparrow(\downarrow)}(E) = -\frac{1}{\pi} \text{Im}[G_{i,i}^{r,\uparrow(\downarrow)}(E)], \tag{8}$$

$$<n_i^{\uparrow(\downarrow)}> = \int_{-\infty}^{+\infty} f(\mu, E) \rho_i^{\uparrow(\downarrow)}(E) \cdot dE, \tag{9}$$

where $f(\mu, E)$ is the Fermi function with the chemical potential $\mu$.

Similar to the band calculations before, we use two sets of matrix equations from Eq. (6). They are coupled to each other by the electron density $<n_i^{\downarrow(\uparrow)}>$ with opposite spins. This is because that in Eq. (6) and (7) $<n_i^{\downarrow(\uparrow)}>$ appears in the Hamiltonians and self-energies, and they also result from $\mathbf{G}^{r,\uparrow(\downarrow)}$ and the integral (Eqs. (8) and (9)). So this is also a self-consistent problem and has to be solved iteratively.

In the NEGF theory, when obtaining the Green's function, the transmission coefficient can be calculated by [28]

$$T^{\uparrow(\downarrow)}(E) = Tr[\mathbf{G}^{r,\uparrow(\downarrow)}(E) \cdot \mathbf{\Gamma}_L^{\uparrow(\downarrow)}(E) \cdot \mathbf{G}^{a,\uparrow(\downarrow)}(E) \cdot \mathbf{\Gamma}_R^{\uparrow(\downarrow)}(E)] \tag{10}$$

where $\mathbf{\Gamma}_\alpha^{\uparrow(\downarrow)}(E) = i[\mathbf{\Sigma}_\alpha^{r,\uparrow(\downarrow)}(E) - \mathbf{\Sigma}_\alpha^{a,\uparrow(\downarrow)}(E)]$ and $\mathbf{G}^{a,\uparrow(\downarrow)}(E) = [\mathbf{G}^{r,\uparrow(\downarrow)}(E)]^+$. If we want to calculate the semi-infinite system, we may set the self-energy of one lead (such as right lead) to be zero, which means $\mathbf{\Sigma}^r = \mathbf{\Sigma}_L^r$. Then the calculated $\mathbf{G}^{r,\uparrow(\downarrow)}$ and $<n_i^{\downarrow(\uparrow)}>$ correspond to a semi-infinite system.

The integral in Eq. (9) can be effective evaluated by the residue theorem [30-31], in which the Fermi function is approximated by the Padé expansion [32]. The details can be seen in the appendix and the reference [33, 37]

## III. RESULT AND DISCUSSION
### A. Periodic systems

In this part we use the band structure calculation to study some ESDW states in the GNR supercells. The tight-binding model with Hubbard term and the density-functional-based-tight-binding (DFTB) model are employed in the calculation.

#### 1. FM-typed and AF-typed ESDW states

In the paper of K.W. Kim and A.K. Kim, a GNR system with two FM orders is proposed, which is induced by two opposite magnetic fields in the leads. These two FM regions are flapped to each other through a single domain wall in the middle.

Here in our SC calculations we further find that there exists such 'FM flap state' in a periodic GNR system. To study the complicated states in the periodic system, a supercell of GNR is used, as shown in Fig. 1 below. With the proper initial condition of $<n_i^{\uparrow(\downarrow)}>$ in the SC calculation as described in Sec. II, we find some stable state with the flapped FM orders in the periodic GNR regions can be reached. Fig. 2 (a1) and (a2) show the spin polarization for this state in a GNR supercell (M=12, N=8). In this state, the GNR edge on the left (or right) side has the same FM order, and this order is flapped into the opposite direction on the other side. From the 3D plot we see that the FM order changes gradually as a sine function. In the middle part the magnetic order is very small, which can be regarded as a domain wall. So this state is a type of ESDW state.

With the similar approach, we also find another SDW excitation in the same GNR supercell (Fig. 2(b1) and (b2)). This state has the AF order: opposite spin polarizations exist on the two edges of GNR, and these AF orders gradually change and flap in another period as an ESDW.

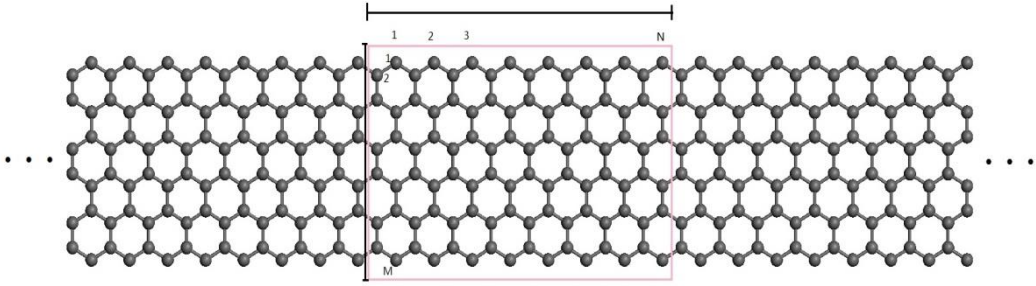

Fig. 1 Schematic description of the zigzag GNR. In the band structure calculation, the rectangle region indicates the unit cell. In the NEGF calculation, the rectangle region indicates the device and the two side regions correspond to the left and right leads. In the unit cell (or the device) there are M atoms in the vertical direction and N atoms on the row of the zigzag edge.

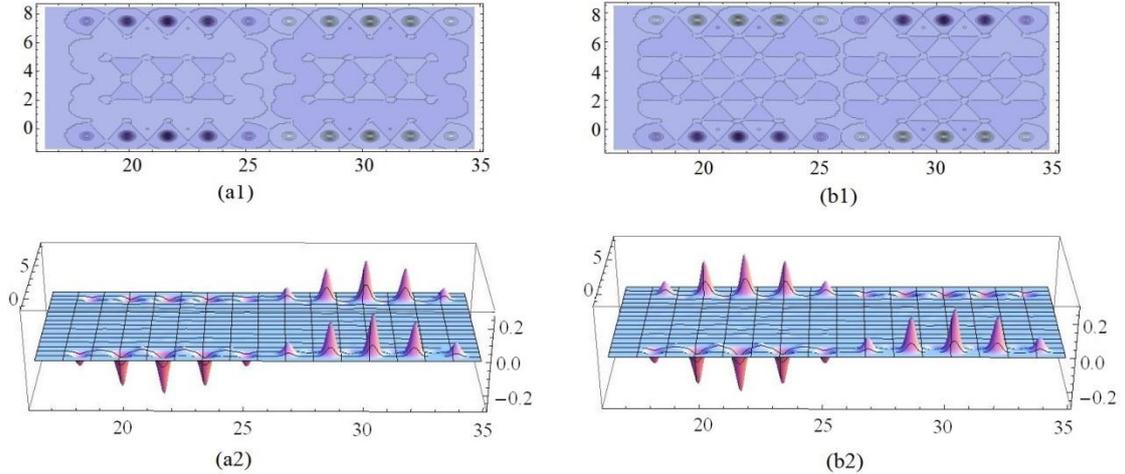

Fig. 2 The ESDW modes with FM order (a1 and a2) and AF order (b1 and b2) in a supercell of zigzag GNR (M=12; N=10). (a1) and (b1) are for the contour plot; (a2) and (b2) are for the 3D plot.

## 2. The first-principles calculations

Now we use the first principle calculation with the DFTB model to confirm these novel spin polarization states. This model is an approximation of density-functional-theory (DFT) method derived from the second-order expansion of Kohn-Sham energy around the reference charge density [35]. A free software package: DFTB+[35-36] developed by the Th. Frauenheim's group is used in our paper. A supercell is used in the band calculations. The vacuum layers of 12 nm in the ribbon plane and 14 nm in the normal direction is applied to avoid the interaction between GNR in different unit cells. In the SCC iteration we choose the Broyden mixing scheme and the energy tolerance is set to 5*10-6 eV.

Fig. 3 (a) shows the calculation results for the FM-typed ESDW state in a GNR supercell (M=12; N=10). To obtain this state, a proper initial spin configuration is set in the input file of DFTB+ package, as well as the TB + Hubbard method before. We see that the DFTB+ software can generate a similar state, compared to that from our TB +Hubbard method (Fig. 2 (a)). The only difference is that there is some shift for the domain wall. But since the system is periodic, it is also reasonable. We also obtain the AF-typed ESDW state by the DFTB+ software, as shown in Fig. 3 (b), which is similar to the state obtained from our TB + Hubbard model (Fig 2 (b)).

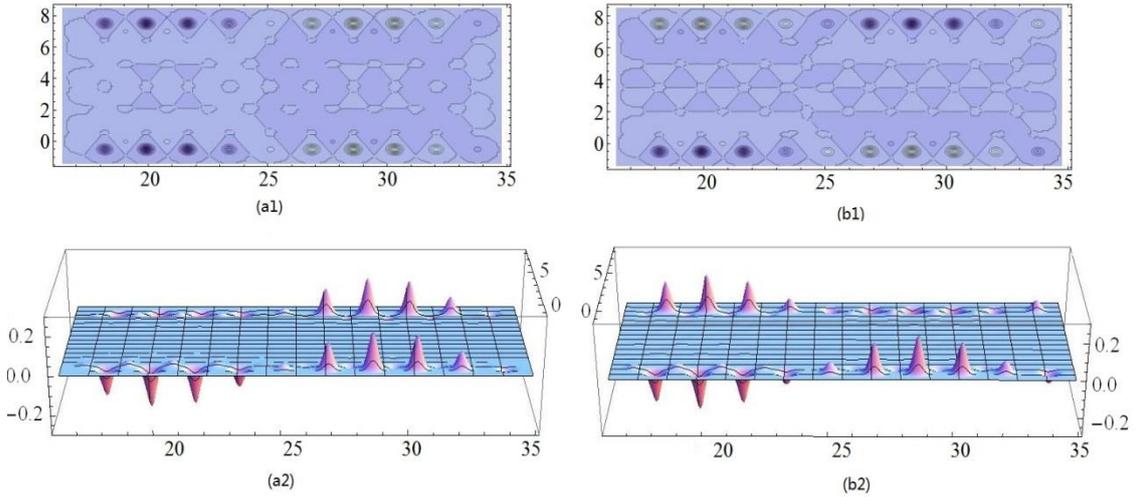

Fig. 3 The ESDW states in a GNR supercell (M=12; N=10) by the DFTB calculation. (a1) and (a2) are the contour/3D plots for the FM order; (b1) and (b2) are the contour/3D plots for the AF order.

## 3. Band structure and Hubbard energy

We also calculated the band structures of these AF/FM-typed ESDW states from the DFTB model and our TB + Hubbard method, as shown in Fig. 4(a) and (b). The system is the same GNR supercell (M=12, N=10). We see that the bands from the DFTB calculation are similar with those from the TB + Hubbard model calculation, except one upper band above the Fermi level. From the results we see that both AF-typed and FM-typed ESDW states have a band gap. This gap comes from the Coulomb repulsion on the zigzag edges as in the pure AF state: The bands near the Fermi level are very flat, which means a large density of states (DOS) near the Fermi level. So the electrons with the same spin on the edge atoms have a large Coulomb repulsion due to the Pauli exclusion principle, which leads the spin separation either in the horizontal or vertical directions.

We observe that the gap (about 0.08 eV) from the DFTB method is smaller than that (0.18 eV)

from the TB + Hubbard model. This is a common fact for these spin excitations in other GNR systems. In the TB calculations we find the gaps in the AF-typed or FM-typed ESDW states are much smaller than that of the pure AF state (0.36eV).

We further investigate the bands near the Fermi level for the two ESDW states and the pure AF state, as shown in Fig. 4(c). Two spin components are degenerate in these bands. The band of the AF state comes from the simple Brillouin zone folding of a spin-polarized GNR system [9]. Here we find that for the FM- and AF-typed ESDW states, one or two gaps are opened at the boundary or center of the Brillouin zone. We explain this as follows. In these ESDW states, two degenerate states in the Fermi surface (here are the Fermi points in 1D case) related by the nesting vector have interaction with each other, which lead a gap at those positions in the reciprocal space.

We also calculate some other GNR supercell systems. The results are shown in Table 1. From the data we see that the gap of the FM-typed ESDW state is usually slightly smaller than that of the AF-typed ESDW state. In a wider GNR supercell, the gap energy difference between the two ESDW states is smaller.

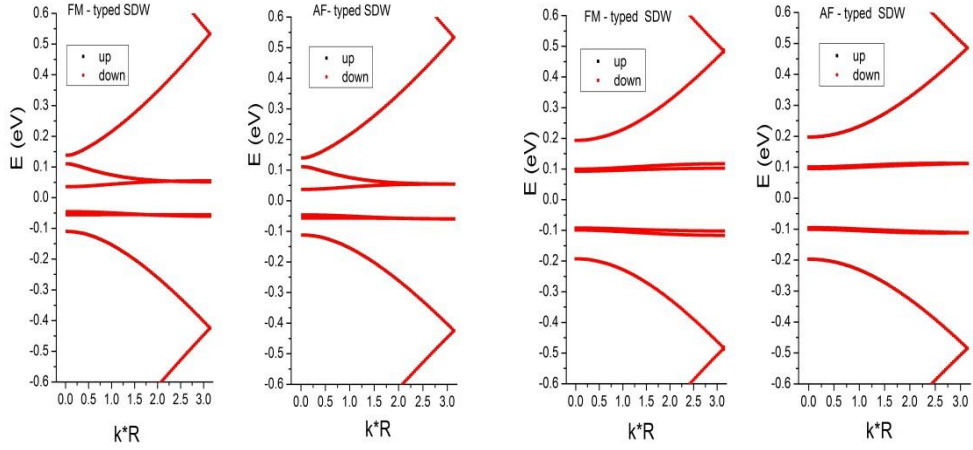

(a)                                                                                       (b)

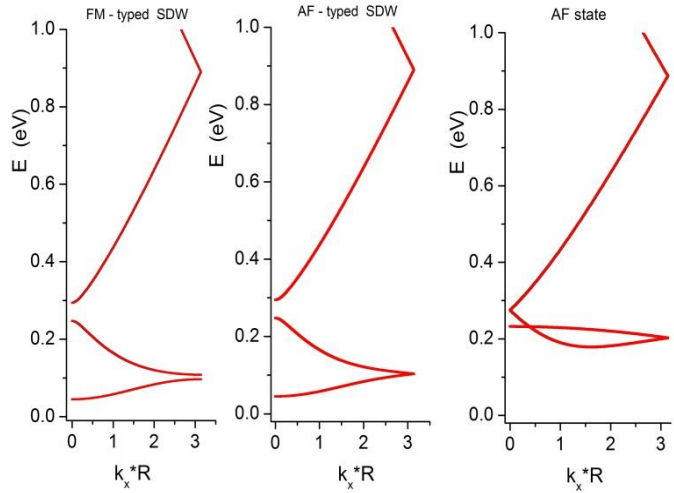

(c)

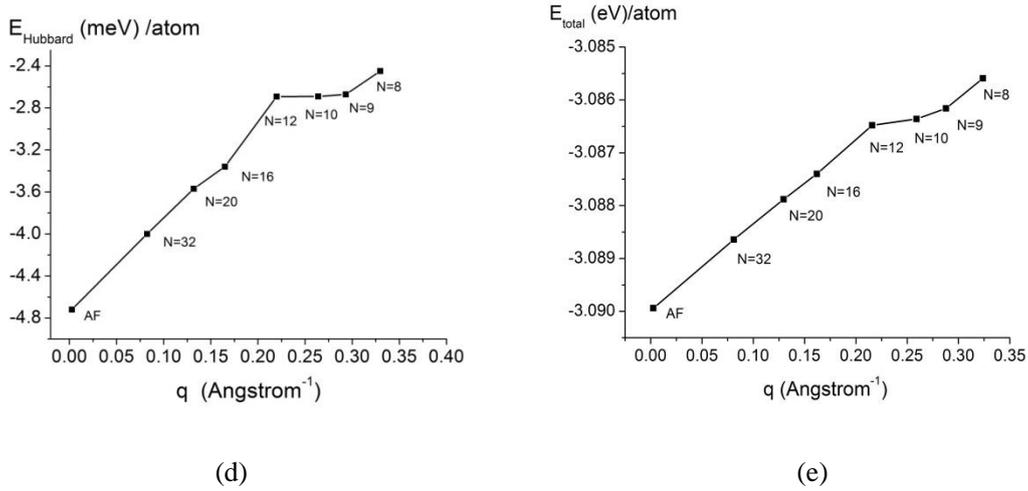

(d)　　　　　　　　　　　　　　　(e)

Fig. 4 (a): The band structures for the ESDW states with FM order (left) and AF order (right) in a GNR supercell (M=12, N=10) with the DFTB model; (b): The band structures for the ESDW states with FM order (left) and AF order (right) in the GNR supercell (M=12, N=10) with the TB + Hubbard model.(c) The bands comparison among the FM-typed SDW state (left), AF-typed ESDW state (middle) and the pure AF state (right) for a GNR supercell (M=12, N=8). (d) The dispersion curve of the AF-typed ESDW excitation; (e) The total energy of the AF-typed ESDW dependence on their wavevectors. In (d) and (e), the width of GNR supercell is set as M=12. And the numbers in the figures denote the period (wavelength) for each ESDW state.

In the TB calculations we find that the involution of the Hubbard term would also result in the change of the kinetic energy. The kinetic energy change is often larger than the Hubbard term's contribution. However, since these novel states come from the spin excitations, here we consider the energies from the Hubbard term. The Hubbard energy ($E_{Hubbard}$) is calculated from the following mean-field Hamiltonian

$$H_U = \sum_i^N (n_{i\uparrow} - \frac{1}{2})(n_{i\downarrow} - \frac{1}{2}) \quad (11)$$

$$H_U \approx U \cdot \sum_i^N [n_{i\uparrow}<n_{i\downarrow}> + n_{i\downarrow}<n_{i\uparrow}> - <n_{i\uparrow}><n_{i\downarrow}> - \frac{1}{4}]. \quad (12)$$

Table 1 shows the band gaps, Hubbard energies and total energies in different ESDW states. The Hubbard energy is averaged in the first Brillouin zone with the Brillouin sampling number $N_k$=400. The large sampling is necessary for a converged result in the SC calculation.

We see that the pure AF state has a lower energy, compared to the pure FM state. This is aggregable with others' work [8,16]. In Table 1 we see that the AF-typed ESDW often has lower energy than the FM-typed ESDW. When the GNR supercell become wider (M changes from 12 to 16), the energy difference between the two ESDW states becomes smaller (with the same supercell length: N=8). As mentioned before, the energy gap difference between two ESDW states in a wider GNR supercell also has such behavior. This means in a wider GNR system the SDW excitations on the two edges have less interaction with each other. In the limit of a very wide GNR, they behave as two independent modes.

We also calculate for other AF-typed ESDW states with different periods (or wavelength) of these supercell systems (M=12). Then we plot their Hubbard energies and total energies versus the wavevectors in Fig. 4 (d) and (e). They are the dispersion curves for these ESDW modes. The dispersion is quite linear near the Brillouin zone center (with very large supercell periods) and then it deviates at large wave vectors. This curve also agrees well with the work in literatures [18, 20]. We find that in some short supercells (for example, M=12, N=7 or less), the AF-typed or FM-typed ESDW states does not exist in our SC calculations. In other words, they are unstable modes. In the paper of F.J. Culchac et al they noticed that some high-energy spin excitation modes in GNR are strongly damped and impossible to observe [19]. Our calculations give similar results to theirs. Here we notice that the ESDW states on the two edges are not the standard harmonic (sine) modes. We see this more apparently in the long supercells as in Fig. 5 later. These eigenmodes often consists of two or more harmonic modes.

| GNR Geometry (M,N) | Magnetization Type | Energy Gap (meV) | Hubbard Energy (meV/atom) | Total Energy (eV/atom) |
|---|---|---|---|---|
| (12,8) | AF order | 357.2 | -4.72 | -3.08994 |
| (12,8) | FM order | ----- | -3.99 | -3.08867 |
| (12,8) | AF-typed ESDW | 89.80 | -2.45 | -3.08559 |
| (12,8) | FM-typed ESDW | 89.22 | -2.43 | -3.08555 |
| (16,8) | AF-typed ESDW | 139.0 | -2.14 | -3.12659 |
| (16,8) | FM-typed ESDW | 138.8 | -2.13 | -3.12658 |

Table 1. The energy gaps, Hubbard and total energies of different ESDW states in different GNR supercells calculated from the TB + Hubbard model.

## 4. Hybrid SDW excitations

With these self-consistent calculations, we also discover a lot of hybrid ESDW states in the GNR supercells. For example, beginning from different initial spin configurations, we obtain the hybrid ESDW state 1 (Fig. 5 (a)), in which the upper edge has a period spin oscillation in the left part (1) and has a flat spin polarization curve in the right part (2). The lower edge has the similar magnetic order but the two parts has exchanged. We also find a similar hybrid ESDW state 2 (Fig. 5 (b)). In this state the upper edge (1 and 3) has a spin oscillation of a large period and the lower edge (2 and 4) has a spin oscillation of a small period. Fig.5 (c) shows the third ESDW state: the upper edge have the pure FM order and the lower edge have a spin flap order (spin wave) with a period of the half GNR length. From this state we can conclude that the two spin orders are almost independent on the upper and lower edges.

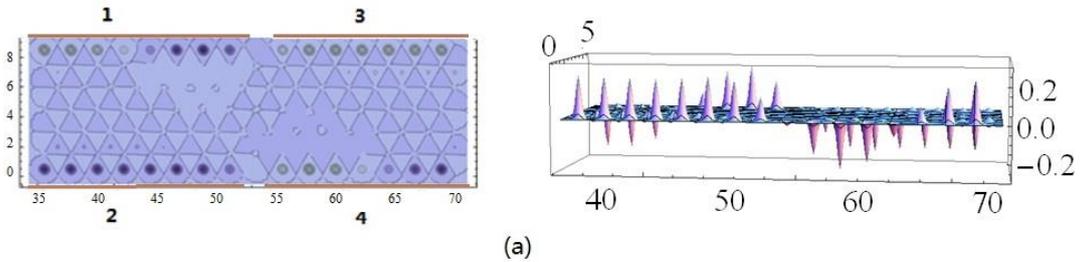

(a)

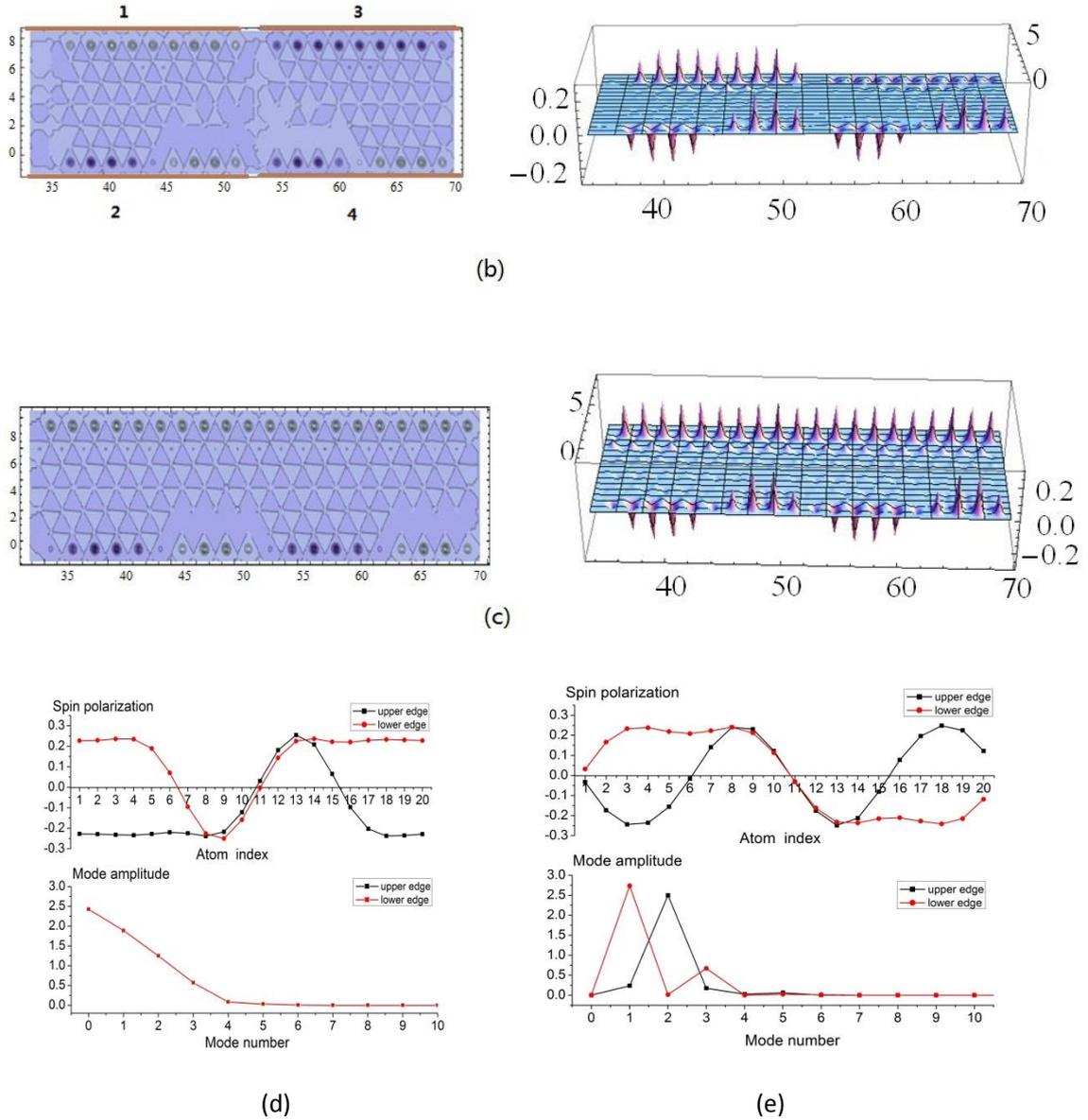

Fig. 5 (a)-(c): Some hybrid ESDW states in GNR supercells. The left part is the contour plot and the right part is the 3D plot for the spin polarization. (a): The hybrid spin flap state 1; (b): The hybrid spin flap state 2; (c): The hybrid spin flap state 3; (d) and (e): The spin polarization on the two ribbon edges (upper part) and the mode analysis (lower part) for the hybrid ESDW states. (d) is for the hybrid state 1 and (e) is for the hybrid state 2. In this figure the GNR supercell's size is set as: M=12, N=20.

We also plot the spin polarizations of the two edges for the above hybrid ESDW state 1 and state 2 as in Fig. 5(d) and (e) (upper parts) respectively. Then we employ a discrete Fourier transformation (DFT) for a harmonic mode analysis of these ESDW states, as shown in Fig. 5(d) and (e) (lower parts). We see that in the hybrid ESDW state 1 there some about 4 modes which contribute to this spin order. In the hybrid ESDW state 2, the harmonic mode 1 (with the period of N) and mode 3 (with the period of N/3) are two main contributions to the lower-edge spin order; and the harmonic mode 2 (with the period of N/2) is mainly attributed to the upper-edge spin order.

## B. Non-periodic systems

Besides the periodic GNR systems, we also use the NEGF theory coupled with the TB + Hubbard model to calculate some non-periodic GNR systems. We find that these interesting spin polarization states also may exist in some local regions of a uniform GNR. The system consists of a device part and two lead parts, as shown in Fig. 1. The calculation process is given in Sec. II.

### 1. Transmission spectra for a FM state

Firstly a benchmark NEGF calculation is done for a uniform infinite GNR (with two leads and a device). We calculate the transmission spectrum (Fig. 6 (a)) for a FM state in such GNR system (M=12, N=8 in the device and M=12, N=2 in the two lead parts). The transmission spectrum agrees well with the FM band structure of a single GNR as shown in Fig. 6(b) (also see Reference [16]). In some energy regions the transmissions are different for two spins in a FM state. The integer transmission plateaus indicate this system is a uniform periodic system.

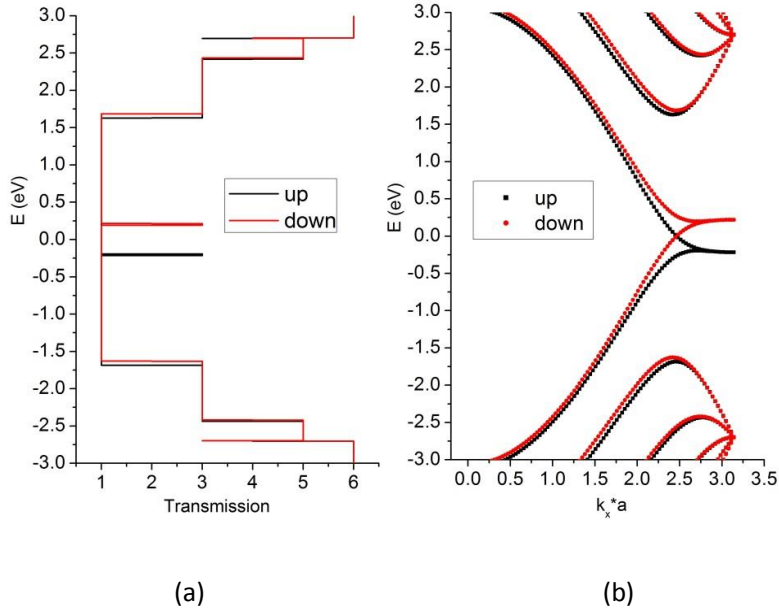

(a) (b)

Fig. 6 (a) The transmission spectrum of a uniform GNR (In the device part: M=12, N=8; in the two lead parts: M=12, N=2) for a pure FM state. (b) The band structure of this FM state in a single GNR.

### 2. ESDW states in lead-device-lead GNR systems

We calculate the Green's function and DOS of the device by using use Eq. (6) - (8) and then do the energy integral with Eq. (9) to obtain the electron density. The contour integral method is used in the integration (see Appendix). After obtaining the electron density, we put them into the Hubbard term and calculate the Green's function again for the iterations until a SC solution is reached. In the beginning of the SC calculation, we set some initial electron densities with two spins in the Hubbard term. So with different initial spin configurations, for example the FM-typed or AF-flap ESDW in the middle of device, are set in the beginning of SC calculation. To calculate the self-energy, electron densities with spin configurations (such as AF order and FM order) are set in the lead Hamiltonians similarly.

Two SC solutions are shown in the following. Fig. 7(a) is the FM-typed SDW excited in the middle of the device, while the leads have the opposite FM states. Fig 7(b) is the AF-typed SDW excited in the middle of the device between while the two leads have the opposite AF states. We also observe that the transmission spectra of these two states (right part of Fig. 7) are almost the same and without any spin separation.

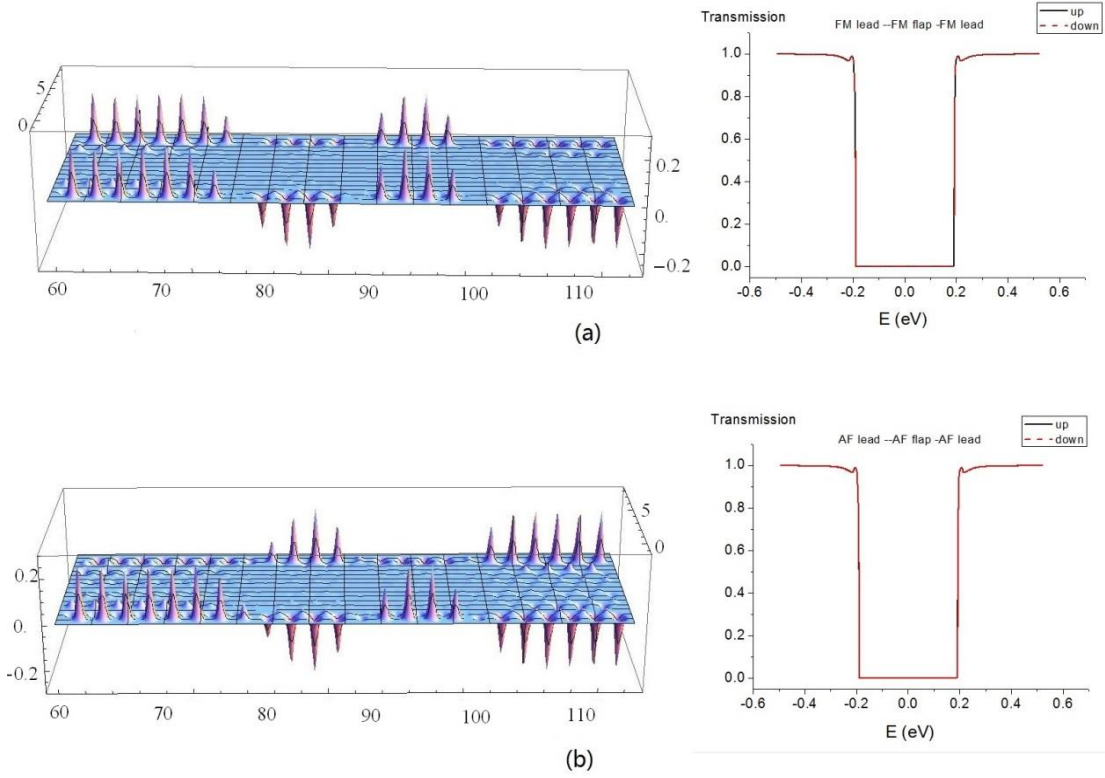

Fig. 7 The spin polarization (left part) and transmission spectra (right part) for the FM-ordered ESDW (a) and AF-ordered ESDW (b) in the lead-device-lead GNR systems. The size of the device part is M=12, N=20; and the size of the lead part is M=12; N=6.

3. ESDW states in semi-infinite GNR systems

Finally we show the excited SDW states in terminal of semi-infinite GNR systems. The calculation process is very similar with that of the lead-device-lead system above, except one lead is not coupled to the device.

Our results are shown in the following. Fig. 8 (a) and (b) are two states with a half period of ESDW (AF or FM order) in the terminal region, and the magnetic orders in the inner region tends to the same (AF or FM order). The polarization near the terminal side is observed to be larger. These states can be explained as the surface ESDW excitations in semi-infinite ribbons from the original AF or FM order.

Similarly, with some other initial spin configurations, we find out another two hybrid ESDW surface states, as shown in Fig. 8 (c) and (d). These two states begin with the FM (or AF) order near the terminal side and then transforms into the AF (or FM) order in the inner region.

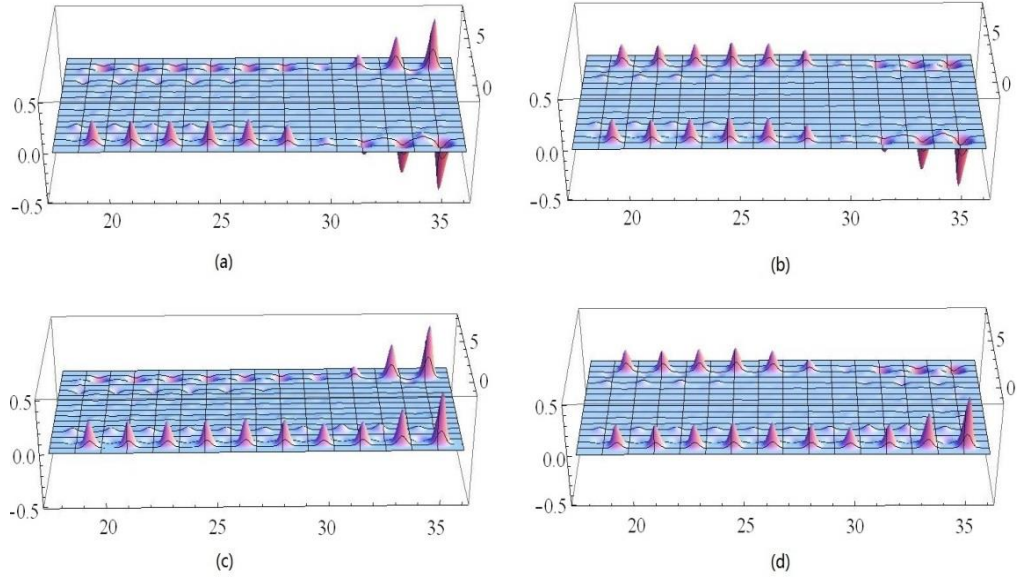

Fig. 8 Some terminal ESDW states in semi-infinite GNR systems (M=12, N=10). (a) AF-ordered ESDW in terminal and AF order in bulk; (b) FM-ordered ESDW in terminal and FM order in bulk; (c) FM-ordered ESDW in terminal and AF order in bulk; (d) AF-ordered ESDW in terminal and FM order in bulk.

## IV. CONCLUSIONS

We have systematically investigated the excited SDW modes in the zigzag GNR. Many interesting spin polarization states have been found, such as the AF-typed and FM-typed ESDW states, and some hybrid ESDW states. The first principle calculation has been employed to confirm these states. We have calculated the Hubbard energy and the band structures of these states. The periods of these ESDW states are commensurate. In the Brillouin zone boundary or center there are some gaps due to the interaction of two degenerate states related by nesting vectors. The dispersion relation of the AF-typed ESDW states has been plotted with a linear relation in the long wavelength limit. In a wider GNR the excited spin polarizations on two edges tend to be independent since the energy difference between the AF- and FM- typed ESDW states is smaller than that in a narrower GNR.

Finally we have utilized the NEGF theory coupled with the Hubbard model to calculate the SDW excitations in non-periodic zigzag GNRs. Some interesting magnetic states such as the middle-part and terminal-part SDW excitation states (with AF/FM order) are found and discussed.

We believe these ESDW are intrinsic spin excitations in a uniform zigzag GNR without any lattice deformation. These ESDW may be excited by some circular polarized light with angular momentum in experiment and detected by some technique such as neutron scattering.

## ACKNOWLEDGEMENTS

We are grateful to Prof. Weiqiang Chen in the Southern University of Science and Technology, China and Dr. Yong Wang in the Nankai University for useful discussions and kind help in the calculation for the Hubbard model. We thank Dr. Rui Wang in the Chongqing University for his kind help in the computer service. Finial support from Chongqing University (Grand No. 0233001104429) and NSFC are also gratefully acknowledged.

# APPENDIX: ELECTRON DENSITY CALCULATION BY THE CONTOUR INTEGRAL METHOD

In this part we give a brief introduction for the contour integral calculation. Eq. (9) is rewritten as

$$<n_i> = \int_{-\infty}^{+\infty} f(\mu, E) \frac{\text{Im}[G^r_{i,i}(E)]}{\pi} \cdot dE . \tag{A1}$$

Here we omit the superscript for spin. We may further write it into the density matrix form as below.

$$\boldsymbol{\sigma} = -\int_{-\infty}^{+\infty} \frac{1}{\pi} f(\mu, E) \text{Im}[\mathbf{G}^r(E)] \cdot dE \tag{A2}$$

It is easy to derive that this integral by the following contour integral [30, 33, 37]. Here we give a brief derivation. We employ the residue theorem to the following contour integral

$$\mathbf{I}_c = \oint f(\mu, E) \mathbf{G}^r(E) dE = \int_{-R}^{R} f(\mu, E) \mathbf{G}^r(E) dE + \int_{C_R} f(\mu, E) \mathbf{G}^r(E) dE \tag{A3}$$

where $C_R$ is the semi-circle with a radius R in the half complex plane.

If we replace the Fermi function with the following Padé approximation [32]

$$f(\mu, E) \approx f_P(E) = \frac{1}{2} + \frac{1}{\beta} \sum_{p=1}^{N_p} \left( \frac{R_p}{E - \mu - z_p^+/\beta} + \frac{R_p}{E - \mu - z_p^-/\beta} \right) \tag{A4}$$

where $R_p$ and $z_p$ is the residue and poles in the Padé approximation. With this approximation, the contour integral is obtained as

$$\mathbf{I}_c = \frac{2}{\beta} \sum_{p=1}^{N_p} R_p \text{Re}[\mathbf{G}^r(z_p)] \tag{A5}$$

Since $\mathbf{G}^r(z)$ has the following limit: $\lim_{z \to \infty} \mathbf{G}^r(z) = \frac{1}{z}$, so the second term in Eq. (A3) can be easily evaluated in the limit of infinite R:

$$\int_{C_R(R \to \infty)} f_P(\mu, E) \mathbf{G}^r(E) dE = \frac{1}{2} \frac{1}{R} \cdot R\pi i = \frac{1}{2} \pi i$$

With some simplification, we finally have

$$\boldsymbol{\sigma} = \frac{\mathbf{I}}{2} + \frac{2}{\beta} \sum_{p=1}^{N_p} R_p \text{Re}[\mathbf{G}^r(z_p)] \tag{A6}$$

where $\mathbf{G}^r(z)$ is the Green's function calculated by the following matrix inverse

$$\mathbf{G}^r(z) = [z\mathbf{I} - \mathbf{H} - \boldsymbol{\Sigma}^r_L(z) - \boldsymbol{\Sigma}^r_R(z)]^{-1} . \tag{A7}$$

Email: xiehangphy@cqu.edu.cn